\documentclass[aps,showpacs,manuscript,12pt]{revtex4}
\usepackage{amssymb}
\usepackage{amsmath}
\usepackage{graphicx}

\begin{document}

\title{\textbf{The computational complexity of traditional
Lattice-Boltzmann methods for incompressible fluids$^{\S }$}}
\author{Marco Tessarotto$^{a,b,c}$, Enrico Fonda$^{d}$
and Massimo Tessarotto$^{c,d}$} \affiliation{\ $^{a}$Civil
Protection Agency, Regione Friuli Venezia-Giulia, Palmanova,
Italy, $^{b}$Department of Electronics, Electrotechnics and
Informatics, University of Trieste, Italy, $^{c}$Consortium of
Magneto-fluid-dynamics, University of Trieste, Italy,
$^{d}$Department of Mathematics and Informatics, University of
Trieste, Italy}

\begin{abstract}
It is well-known that in fluid dynamics an alternative to
customary direct solution methods (based on the discretization of
the fluid fields) is provided by so-called \emph{particle
simulation methods}. Particle simulation methods rely typically on
appropriate \emph{kinetic models} for the fluid equations which
permit the evaluation of the fluid fields in terms of suitable
expectation values (or \emph{momenta}) of the kinetic distribution
function $f(\mathbf{r,v},t),$ being respectively $\mathbf{r}$
and\textbf{\ }$\mathbf{v}$ the position an velocity of a test
particle with probability density $f(\mathbf{r,v},t)$. These
kinetic models can be continuous or discrete in phase space,
yielding respectively \emph{continuous} or \emph{discrete kinetic
models} for the fluids. However, also particle simulation methods
may be biased by an undesirable computational complexity. In
particular, a fundamental issue is to estimate the algorithmic
complexity of numerical simulations based on traditional LBM's
(Lattice-Boltzmann methods; for review see Succi, 2001
\cite{Succi}). These methods, based on a discrete kinetic
approach, represent currently an interesting alternative to direct
solution methods. Here we intend to prove that for incompressible
fluids fluids LBM's may present a high complexity. The goal of the
investigation is to present a detailed account of the origin of
the various complexity sources appearing in customary LBM's. The
result is relevant to establish possible strategies for improving
the numerical efficiency of existing numerical methods.
\end{abstract}

\date{\today }
\maketitle



\section{Introduction: basic motivations}

Fluid Dynamics represents, is a sense, one of the unsolved
problems of contemporary science. In fact, although \emph{fluid
equations,} i.e., the complete set of differential equations which
are associated to a fluid are in many cases well known (and
therefore to be considered as\emph{\ phenomenological equations}),
a complete knowledge of their solutions is not
achievable. This justifies the constant efforts placed in \emph{%
Computational Fluid} \emph{Dynamics} (CFD) to develop more
efficient numerical simulation methods, particularly suitable for
software implementation on parallel supercomputers\ and able to
simulate the small-scale dynamics of complex fluid systems. This
explains the increasing role of CFD both for theoretical research
and applied sciences, in particular for the development of
industrial applications. CFD has the goal
of determining the numerical solutions of suitable fluid equations,\emph{\ }%
associated to a prescribed fluid,\emph{\ }which are
expressed\emph{\ }in
terms of appropriate physical observables, the so-called \emph{fluid fields}%
. For neutral isothermal fluids the fluid equations are the \emph{%
incompressible Navier-Stokes equations} (\emph{INSE}), which are
represented respectively by the Navier-Stokes equation, advancing
in time the fluid velocity $\mathbf{V}(\mathbf{r},t),$ the Poisson
equation which, at a given time, determines the fluid pressure
$p(\mathbf{r},t)$ and the isochoricity condition which forces the
fluid velocity to result everywhere divergence-less.

A basic aspect of numerical algorithms is their
\textsl{computational complexity}, namely the number of
(elementary) discrete logical operations which must be performed
in a given time interval to carry out a prescribed calculation. In
the case of CFD this is measured by the number of discrete
operations required to advance in time the fluid fields in a
prescribed finite time interval. The computational complexity of
numerical solution methods for INSE depends obviously both on the
"physical" properties of the fluid, i.e., on the characteristic
spatial and time scales of the fluid fields ($L,T$)$,$ as well as
on the the accuracy with which the solutions are actually
determined. In particular, the numerical methods used in CFD and
CMFD for incompressible fluids (see for example J. Kim, 1999
\cite{Kim99}),
exhibit typically a high \textsl{%
computational} (or \textsl{algorithmic}) \textsl{complexity},
although due to different reasons. Among the most popular are the
so-called \textsl{direct solution methods}, which are\ based on
the discretization of the fluid fields (see, for example J. Kim
and P. Moin, 1979 and 1985 \cite{Kim79,Kim85}; Z. Warsi,1995
\cite{rey}; J. Ferzinger and M. Peric, 1996 \cite{Ferziger}; S.
Turek, 1999 \cite{Turek}). Especially for LES (large eddy
simulations) of fluid problems with complex boundaries and in the
presence of strong fluid turbulence these numerical methods are
difficult to implement on supercomputers based on parallel
architectures, mostly due to the difficulty of solving the fluid
equation which advances on time the fluid pressure (i.e., the
Poisson equation). In the worst cases (i.e., when the turbulence
scale length becomes comparable to the size of the spatial cells
and is sufficiently widespread in the domain of the fluid), the
latter exhibits a computational complexity proportional to
$N^{\gamma },$ where the exponent $\gamma $ is typically larger
than $1$ and in the worst case can be $\gamma =2$ and $N$ \ is the
number of nodes (or sets) in which the fluid is discretized (and
which measures the \textquotedblright size\textquotedblright\ the
numerical simulation). This phenomenon may result, in actual
numerical experiments on
incompressible fluids, in the appearance of a\emph{\ computational bottleneck%
} which effectively limits the size of numerical simulations. \
Since all fluids with sufficiently small Mach number behave as
incompressible, it is obvious that this phenomenon represents a
major issue in computational fluid dynamics and a formidable
obstacle to the development of large scale numerical simulations
relevant for actual industrial applications.

A key aspect of CFD lies also in the search of efficient
\textsl{parallel numerical algorithms}, namely algorithms which
can be represented in terms of independent or weakly related
sub-algorithm to be handled independently by\ a prescribed set of
processors working in parallel. It is well known that customary,
most algorithmically-efficient, direct-solution methods adopted in
CFD, become highly inefficient for parallel processing. This is
due to the Poisson equation, an elliptic PDE which results
difficult to solve numerically (the best numerical methods turn
out to be the least efficient for parallel processing). This has
motivated the search of numerical algorithms which are able to
determine the fluid pressure without actually solving numerically
the Poisson equation.

\section{Lattice-Boltzmann methods}

The investigation of lattice Boltzmann methods (LBM) has drawn
considerable attention in the last few years (for a review see
\cite{Succi2002}). The simplicity of the algorithms, along with
the ability to determine the fluid pressure without actually
requiring the explicit numerical solution of the Poisson equation,
have allowed LBM's to emerge as possible alternatives to
traditional CFD approaches based on the direct discretization of
the relevant fluid equations (direct solution methods). A
specialized interesting area of investigation concerns, in
particular, the\ hardware implementation of LB algorithms for
reconfigurable computing \cite{Marco
Tessarotto2003,Tessarotto2004}, a technique which appears
especially relevant for applications to real-time simulations and
process optimization in industrial fluid dynamics. In addition,\
these approaches exhibit a natural \textsl{parallelism} and - at
the same time - result highly \textsl{scalable }on an array of
independent processing units working in parallel. In essence, this
is because the elementary process (represented by the evolution of
the fluid fields) is reduced to the time evolution of \ a suitable
set of weakly correlated test particles, each one represented by a
discrete kinetic distribution function. In customary LBM's
typically the fluid is treated as weakly compressible, with
asymptotically small deviations from the condition of true
incompressibility. This is realized by actually replacing INSE
with a modified set of fluid equations. In particular the pressure
is determined in this case by an equation of state, while the
fluid velocity is advanced in time by means of the kinetic
distribution function. This permits to avoid the computational
complexity of Poisson equation and to obtain, at the same time, a
highly parallelizable approximate solution method. For this reason
LB approaches are currently considered as a promising
computational tool for massive fluid-dynamics numerical
simulations in a wide spectrum of applications, ranging from
isothermal incompressible fluids, to thermal, compressible fluids,
to MHD dynamics in fusion plasmas.

\section{Difficulties with LBM's}

Despite the significant number of theoretical and numerical papers
appeared in the literature in the last few years, the lattice Boltzmann method \cite%
{McNamara1988,Higueras1989,Succi1991,ChenpChen-1991,Benzi1992,Chen1992}
- among many others available in CFD\ - is probably the one for
which a complete understanding is not yet available. Although
originated as an extension of the lattice gas automaton
\cite{Frisch1986,Frisch1987} or a special discrete form of the
Boltzmann equation \cite{He1997}, several aspects regarding the
very foundation of LB theory still remain to be clarified.
Consequently, also the comparisons and exact relationship between
the various lattice Boltzmann methods (LBM) and other CFD methods
are made difficult or, at least, not yet well understood. Needless
to say, these comparisons are essential to assess the relative
value (based on the characteristic computational complexity,
accuracy and stability) of LBM and other CFD methods. In
particular the relative performance of the numerical methods
depend strongly on the characteristic spatial and time
discretization scales, i.e., the minimal spatial and time scale
lengths required by each numerical method to achieve a prescribed
accuracy. On the other hand, most of the existing knowledge of the
LBM's properties originates from numerical benchmarks (see for example \cite%
{Martinez1994,Hou1995,He1997b}). Although these studies have
demonstrated the LBM's accuracy in simulating fluid flows, few
comparisons are available on the relative computational efficiency
of the LBM and other CFD methods \cite{He1997,He2002}. \ The main
reason [of these difficulties] is probably because current LBM's,
rather than being exact Navier-Stokes solvers, are at most
asymptotic ones (\emph{asymptotic LBM's}), i.e., \ they depend on
one or more infinitesimal parameters and recover INSE only in an
approximate asymptotic sense. The motivations of this work are
related to some of the basic features of customary LB theory
representing, at the same time, assets and weaknesses. One of the
main reasons of the popularity of the LB approach lays in its
simplicity and in the fact that it provides an approximate Poisson
solver, i.e., it permits to advance in time the fluid fields
without explicitly solving numerically the Poisson equation for
the fluid pressure. However customary LB approaches can yield, at
most, only asymptotic approximations for the fluid fields. This is
because of two different reasons. The first one is the difficulty
in the precise definition of the kinetic boundary conditions in
customary LBM's, since sufficiently close to the boundary the form
of the distribution function prescribed by the boundary conditions
is not generally consistent with hydrodynamic equations. The
second reason is that the kinetic description adopted implies
either the introduction of weak
compressibility \cite%
{McNamara1988,Higueras1989,Succi,Benzi1992,ChenpChen-1991,Chen1992}
or temperature \cite{Ansumali2002} effects of the fluid or some
sort of state equation for the fluid pressure \cite{Shi2006}.
These assumptions, although physically plausible, appear
unacceptable from the mathematical viewpoint since they represent
a breaking of the exact fluid equations.  Moreover, in the case of
very small fluid viscosity customary LBM's may become inefficient
as a consequence of the low-order approximations usually adopted
and the possible presence of numerical instabilities (see below).
These accuracy limitations at low viscosities can usually be
overcome only by imposing severe grid refinements and strong
reductions of the size of the time step. \ This has the inevitable
consequence of raising significantly the level of computational
complexity in customary LBM's (potentially much higher than that
of so-called direct solution methods), which makes them
inefficient or even potentially unsuitable for large-scale
simulations in fluids. A fundamental issue is, therefore, related
to the construction of more accurate, or higher-order,
LBM's{\small ,} applicable for arbitrary values of the relevant
physical (and asymptotic) parameters. However, the route which
should permit to determine them is still uncertain, since the very
existence of an underlying exact (and non-asymptotic)
discrete kinetic theory, analogous to the continuous inverse kinetic theory \cite%
{Ellero2004,Ellero2005}, is not yet known. According to some authors \cite%
{Shan1998,Ansumali12002,Chikatamarla2006} this should be linked to
the discretization of the Boltzmann equation, or to the possible
introduction of weakly compressible and thermal flow models.
However, the first approach is not only extremely hard to
implement \cite{Bardow}, since it is based on the adoption of
higher-order Gauss-Hermite quadratures (linked to the
discretization of the Boltzmann equation), but its truncations
yield at most asymptotic theories. Other approaches, which are
based on 'ad hoc' modifications of the fluid equations (for
example, introducing compressibility and/or temperature effects
\cite{Ansumali2005}), by definition cannot provide exact
Navier-Stokes solvers. Another critical issue is related to the numerical stability of LBM's \cite%
{Succi2002},{\small \ }usually attributed to the violation of the
condition of strict positivity (\emph{realizability condition})
for the kinetic distribution function
\cite{Boghosian2001,Succi2002}. Therefore, according to this
viewpoint, a stability criterion should be achieved by imposing
the existence of an H-theorem (for a review see
\cite{McCracken2005}). In an effort to improve the efficiency of
LBM numerical implementations and to cure these instabilities,
there has been recently a renewed interest in the LB theory.
Several approaches have been proposed. The first one involves the
adoption of entropic LBM's (ELBM \cite%
{Karlin1998,Karlin1998aa0,Karlin1999,Boghosian2001} in which the
equilibrium distribution satisfies also a maximum principle,
defined with respect to a suitably defined entropy functional.
However, usually these methods lead to non-polynomial equilibrium
distribution functions which potentially result in higher
computational complexity and lower numerical accuracy
\cite{Yong2003}. Other approaches rely on the adoption of multiple
relaxation times. However the efficiency, of these methods is
still in doubt. Therefore, the search for new [LB] models,
overcoming these limitations, remains an important unsolved task.

\section{Asymptotic LBM's : computational complexity}

An alternative to direct solution methods, which can reduce
significantly the complexity caused by Poisson equation, may be
achieved by so-called \emph{particle simulation methods,} in which
the dynamics of fluids is approximated in terms of a set of
\emph{test particles }which advance in time in terms of suitable
evolution equations\emph{\ defined in such a way to satisfy
identically the Poisson} \emph{equation}. Particle simulation
methods rely typically on appropriate \emph{kinetic models} for
the fluid (or magnetofluid) equations which permit the evaluation
of the fluid fields in terms of suitable expectation values (or
\emph{momenta}) of the kinetic distribution function
$f(\mathbf{r,v},t),$ being respectively $\mathbf{r}$ and\textbf{\
}$\mathbf{v}$ the position an velocity of a test particle with
probability density $f(\mathbf{r,v},t)$. These kinetic models can
be continuous or discrete in phase space, yielding respectively \emph{continuous%
} or \emph{discrete kinetic models} for the fluids. In particular,
discrete models are those in which the kinetic distribution
function is discretized in some sense (such as configuration space
and/or velocity space and/or time), which usually leads to the
approximate description of a continuum (the fluid) in terms of a
discrete and finite set of test particles, each described by a
suitable distribution function $f_{i}$.  An example is provided
by \emph{LBM}'s (\emph{Lattice-Boltzmann methods})%
\emph{\ }in which the distribution function $f_{i}$ is discretized
in velocity space, i.e., is defined only for a suitable set of
discrete velocities \ $\mathbf{a}_{i}$ ($i=0,n$) \ - to be
identified with the velocities of test particles - which are all
assumed as constant in time. However, particle simulation methods
- and in particular LBM's - may exhibit, in their turn, different
sources of undesirable algorithmic complexity. In particular,
so-called \emph{asymptotic kinetic theories,} i.e. kinetic
theories which recover the prescribed set of fluid equations only
in suitable asymptotic limits, will depend necessarily from
dimensionless and positive parameters $\varepsilon
_{1},..,\varepsilon _{k},$ all assumed $\ll 1,$ and demand
typically that the kinetic distribution function be determined
accurate to prescribed order in \ $\varepsilon _{1},..,\varepsilon
_{k}.$ For example, the parameters $\varepsilon
_{1},..,\varepsilon _{k}$ my be identified, in some cases, with
$1/N_{k}$, being $N_{K}$ the so-called Knudsen number$,$ $N_{M}$
the Mach number, etc.-  Customary LBM's  \cite
{McNamara1988,Higueras1989,Succi1991,ChenpChen-1991,Benzi1992,Chen1992}
are typically constructed in such a way to satisfy the exact fluid
equations (INSE) only in an asymptotic sense and \emph{are
therefore asymptotic.} These requirements inevitably give rise to
additional computational complexity in numerical simulations based
on asymptotic kinetic theories \emph{which are nevertheless free
of the Poisson equation complexity described above}. It is
interesting to mention the possible sources of algorithmic
complexity affecting customary LB methods, which are related to
the choices of the time and space discretization scale lengths
$\Delta L$ and $\Delta t$ adopted in customary LBM's (which
determine the number of cell and elementary time intervals in
which the fluid domain and the time interval are divided). They
are all essentially a consequence of the fact that these numerical
schemes are based on asymptotic kinetic theories, i.e., they are
characterized by an \emph{asymptotic parameter} $\varepsilon ,$ to
be assumed non negative and $\ll 1.$ The parameter $\varepsilon $
enters the theory through the discrete kinetic distribution
function $f_{i},$ which describes the time evolution of the $i-$th
test particle in a given position (node)$.$ In particular in order
that the kinetic theory recovers the correct fluid equations the
kinetic distribution function must remain at all times suitably
close to an appropriate \textquotedblright
equilibrium\textquotedblright\ distribution function
$f_{i}^{(eq)}$, in the sense that denoting $\delta f_{i}\equiv
f_{i}-f_{i}^{eq}$ the \textquotedblright
deviation\textquotedblright\ from the equilibrium distribution,
there must result
\begin{equation}
\delta f_{i}\sim o(\varepsilon )  \label{ord-0}
\end{equation}%
(\textsl{relaxation condition for the kinetic distribution
function}). In asymptotic LBM's this condition is usually
satisfied by adopting an LB-BGK
kinetic equation (a kinetic equation with a BGK collision operator \cite%
{BGK1954}) which is characterized by a relaxation time $\tau >0$.
On the other hand, since these methods are all based on the Euler
approximation for the streaming operator, a basic consequence [of
these assumptions] is that the amplitude of the time step ($\Delta
t$) used to advance in time the
kinetic distribution function $f_{i}$ must result such that%
\begin{equation}
\frac{\Delta t}{\tau }\sim o(\varepsilon )\ll 1.  \label{ord-1a}
\end{equation}%
In customary LBM's the parameter $\tau $ is usually linearly
proportional to
the kinematic viscosity of the fluid ($\nu $). For example for the 9Q2D-($%
p-V $)-LB scheme there results $\nu =c^{2}\tau /3.$ Hence it follows%
\begin{equation}
\frac{3c^{2}\Delta t}{\nu }\sim o(\varepsilon )\ll 1.
\end{equation}%
($\nu -$ \textsl{complexity of LBM's}).It follows that $\Delta t$
can become very small for weakly viscous fluids, with the
consequence of \ increasing significantly the computational
complexity of asymptotic LBM's (with respect to direct solution
methods).

Another potential source of complexity is given by the requirement
that the fluid fields must be suitably smooth and slowly varying
on the relevant discretization scales, i.e., in particular such
that they satisfy the asymptotic ordering
\begin{equation}
\frac{\Delta L}{L},\frac{\Delta t}{T}\sim \varepsilon
\end{equation}%
(\textsl{smallness of discretization scales}), being $\Delta L$
and $\Delta t,$ respectively, the size of the spatial cells in
which the fluid is discretized and the time step for advancing in
time the discretized kinetic distribution function, while $L$ and
$T$ are the characteristic length and time scales of the fluid
fields. Due to the low-order approximations used in customary LB
methods these are expected to reproduce the correct fluid
equations only in an average sense on the relevant characteristic
scales $L$ and $T$ and not pointwise, i.e., in each node in which
the fluid is discretized. This justifies, for example, the
introduction of so-called \textquotedblright half-way bounce
back\textquotedblright\ boundary conditions in which the
conditions of incompressibility and no-slip at a fixed wall in
contact with the fluid are satisfied only in an average sense in
the boundary cells. Finally, a further source of complexity for
customary LB methods is that they satisfy the condition of
incompressibility only when a suitably defined effective Mach
number, $M^{eff},$ results small enough, which in this case must
be at least of order
\begin{equation}
M^{eff}\sim o(\varepsilon ^{\alpha })\ll 1,  \label{ord-1}
\end{equation}%
where typically $\alpha =3/2$ (\emph{small effective Mach number assumption}%
). In literature the parameter $M^{eff}$ is usually identified
with the ratio between the magnitude of the flow velocity $V$ and
the constant test particle velocity $c:$
\begin{equation}
M^{eff}=\frac{V}{c},
\end{equation}%
which means that the velocity of test particles must be much
larger than the flow velocity, a fact which, by itself, produces a
significant source of computational complexity
($M^{eff}-$\textsl{complexity of LBM's}). A further consequence is
that the \textsl{Courant number} ($N_{C}\equiv \frac{V\Delta
t}{L},$ being $L$ the magnitude of the local scale length used for
the spatial discretization of the fluid domain) of asymptotic
LBM's
results of order%
\begin{equation}
N_{C}\sim o(\varepsilon ^{\beta })\ll 1,  \label{ord-1.1}
\end{equation}
with $\beta \geq \alpha $. As a consequence the amplitude of the time step ($%
\Delta t$) used in customary asymptotic LBM's to advance in time
the fluid
fields results of order%
\begin{equation}
\Delta t\sim M^{eff}\frac{\Delta L}{L}\Delta t_{Opt} \label{Eq.4b}
\end{equation}
($N_{C}-$\textsl{complexity of LBM's}), where $\Delta t_{Opt}$ is
the time step adopted by \emph{optimized numerical solution
methods} in CFD (such as those based on spectral methods), for
which the Courant number $N_{C}$ results of order unity, while
$\Delta L$ and $L$ denote respectively the local grid size adopted
by the LBM and the local characteristic scale length of the fluid
fields. Despite the absence of the N-complexity for them,
asymptotic LBM's may result potentially (much) slower than
optimized numerical methods (such as spectral methods). In fact,
as indicated above they typically require both $M^{eff}\ll 1$ and
$\frac{\Delta L}{L}\ll ,$ namely $\Delta t\ll \Delta t_{Opt}.$

\section{Conclusions}

A fundamental issue in CFD\ is therefore the search of algorithms
with \textsl{reduced algorithmic complexity }and at the same time
with\textsl{\ improved algorithmic efficiency}. This problem has
been investigated in the framework of so-called \emph{inverse
kinetic approaches}, i.e., kinetic theories which are able to
provide, with prescribed accuracy, all fluid equations expressed
as suitable moment equations of the relevant kinetic
equation, to be assumed either continuous \cite%
{Tessarotto2005A,Tessarotto2005B} or discrete \cite{Fonda2007}. As
indicated above, the CMFD Consortium has promoted and developed in
the last few years an intense research activity. Part of the
research activity has concerned inverse kinetic theories for
Newtonian isothermal and non-isothermal fluids. In particular,
first, continuous, non asymptotic, inverse kinetic theories for
incompressible Newtonian fluids have been investigated, with the
goal of developing optimal algorithms which do not exhibit the feature of the $N-$%
\emph{complexity }and, at the same time, do not require subsidiary
asymptotic conditions to be satisfied, such as the requirement of
\textquotedblright closeness\textquotedblright\ in some sense to a
suitable kinetic equilibrium. In particular, a basic features of
the inverse kinetic theory developed is that only mild
restrictions must be satisfied by the kinetic distribution
function in order to satisfy the relevant fluid equations, while
the fluid pressure is advanced in time without solving explicitly
Poisson's equation for the fluid pressure. An interesting issue is
the possibility of formulating for incompressible Newtonian fluids
a discrete kinetic approach, based on a Lattice-Boltzmann scheme
which avoids or reduces the computational bottlenecks
characteristic of previous numerical solution methods of this type
and, at the same time, yield with prescribed accuracy the
solutions of the fluid equations. This includes the basic feature
of determining the fluid pressure without solving explicitly the
Poisson equation.

\section*{Acknowledgments}
Work developed in cooperation with the CMFD Team, Consortium for
Magneto-fluid-dynamics (Trieste University, Trieste, Italy). \
Research developed in the framework of the MIUR (Italian Ministry
of University and Research) PRIN Programme: \textit{Modelli della
teoria cinetica matematica nello studio dei sistemi complessi
nelle scienze applicate}. The support COST Action P17 (EPM,
\textit{Electromagnetic Processing of Materials}) and GNFM
(National Group of Mathematical Physics) of INDAM (Italian
National Institute for Advanced Mathematics) is acknowledged.

\section*{Notice}
$^{\S }$ contributed paper at RGD26 (Kyoto, Japan, July 2008).
\newpage

\bigskip

\end{document}